# Pressure and high $T_c$ superconductivity: applications to sulfur hydrides

Lev P. Gor'kov [1, 2] and Vladimir Z. Kresin [3]

[1]NHMFL, Florida State University, 1800 E. Paul Dirac Drive, Tallahassee, Florida, 32310, USA
[2]L.D. Landau Institute for Theoretical Physics of the RAS, Chernogolovka, 142432, Russia
[3]Lawrence Berkeley Laboratory, University of California, 1 Cyclotron Road, Berkeley, CA 94720, USA

(Dated: ) PACS numbers: 74.20. -z, 74.62.Fj, 63.20.kd, 71.15. -m

The rapid variation of the superconductivity $T_C$ in hydride sulfur (H$_3$S) under high pressure [A. Drozdov et al, Nature **525**, 73 (2015); M. Einaga et al, arXiv: 1509.03156] in a vicinity $P_{cr} \approx 123 GPa$ is interpreted in terms of the 1$^{st}$ order transition, possibly related to a CDW-instability with a non-zero structural vector. The superconductivity mechanisms in high $T_C$ phase be discussed, ordinary methods of calculating $T_C$ are shown not applicable in H$_3$S because, beside the acoustic branches, its phonon spectrum contains hydrogen modes with much higher frequency. A modified approach provides realistic $T_C$ values. The isotope effect (change of $T_C$ at the substitution of deuterium for hydrogen) owes its origin to the high frequency phonons and is different in different phases. The decrease of $T_C$ after reaching a maximum in high-$T_C$ phase is due to the interaction with the second gap arising on a hole-like pocket. The phonon-mediated pairing on the latter is in a non-adiabatic regime of the frequency $\omega_0$ of hydrogen modes comparable to the Fermi energy.

*Introduction.* Search for high-$T_c$ ("room") superconductivity at high pressure was triggered by the suggestion [1] that as in the BCS theory the temperature $T_c$ of transition is proportional to the frequency of phonons mediating the pairing, higher $T_c$ expected in systems comprised of light atoms, including metallic hydrogen. Half a century since superconductivity at 190 K was claimed in sulfur hydrides under pressure $P > 150 GPa$ [2]. Recently, $T_c = 203K$ was confirmed in H$_3$S formed in the decomposition of H$_2$S under pressure [3].

This presentation mainly concerns with the pressure dependence of the superconducting temperature in sulfur hydride H$_3$S. In recent data [4] $T_c \approx 100K$ at $P_{cr} \approx 123GPa$ sharply increases to $T_c \approx 200K$ at $P \approx 150GPa$ as in **a** phase transition. Notably, once $T_C$ reaches its maximum value $T_c \approx 200K$ at the onset of high-$T_C$ phase it immediately *decreases* with the further pressure application [3, 4].

We argue that the 1$^{st}$ order phase transition is a most credible hypothesis for almost doubling of $T_C$ in the narrow experimental pressure interval $\Delta P \approx 25GPa$ and discuss factors that apparently account for so significant $T_C$-increase. Coupling between the superconducting order parameters on hole-like pockets and on large Fermi surfaces is the microscopic mechanism underlying the subsequent $T_c$ decrease.

Before proceeding, one must realize the difficulties that confront the theory in the pressure regime [2-4]. The metallization and superconductivity in sulfur hydrides occur when their basic physical properties are unknown. The theory had to establish the stoichiometry, to prove the stability of the phases emerging at metallization, identify transformations with pressure between the phases, to reconstruct the electronic bands and the phonons spectra, all from the *first principles*. Having the last data, one may attempt to evaluate the temperature $T_c$ of superconducting transition. Such a program was realized in [5-22] having predicted in [5,6] successfully the H$_3$S-stoichiometry, in agreement with the *X*-ray experiment [4]. Nevertheless, one finds inconsistencies between different theoretical works especially apparent in uncertainty for the phase transitions points and the phases' symmetry. Besides, among factors that may lead to the first order transition can also be the CDW instability not considered yet in the literature.

The mechanism of superconductivity in the high-$T_c$ phase in most publications is the phonon mediated electron-electron pairing on *large* Fermi surfaces. With the electron and the phonon spectrum known, the transition temperature $T_c$ is found numerically from the Migdal-Eliashberg (ME) equations [23,24]. Most of these algorithms were elaborated for the ordinary metals. Applicability of the same methods to calculating the superconducting transition temperature for H$_3$S is scrutinized in the next Section; a few drawbacks in [5,6,14-17,20] at solving ME equations are pointed out.

The isotopic dependence of $T_C$ (i.e., its change at the substitution of deuterium for hydrogen) turns out be different on the two sides of the phase transition, in agreement with the experiments [3, 4]. We infer that the key role in superconductivity of H$_3$S [2-4] belongs to high frequency hydrogen modes.

The superconductivity mechanisms discussed in [5-16,20] assert the prevailing role of the Cooper pairing on large Fermi surfaces. Standing apart is a scenario [17-19], in which superconductivity in high-$T_C$ phase is driven by the pairing on small hole-like pockets emerging at several spots of the Brillouin zone (BZ) via the Lifshitz 2.5-topological transition [25,26].

Hole-like pockets in the band structure of the high-$T_c$ phase were theoretically disclosed in [6, 7, 9,11,17,22]. Their especial role in [17] owes to a van Hove (*vH*) singularity peak in the density of states (DOS) in a close vicinity of the chemical potential enhancing strongly the electron-phonon interactions. A peak in DOS is present in several band structure calculations [6,11, 12(Fig.4, Suppl. Mat.), 17-19], but it lies $0.17 \div 04 eV$ under the chemical potential. The results below better agree with the idea that profiling is the interactions at large Fermi surfaces with pockets playing only a supportive role.

Experiment [4] and the theory [5-17] agree upon the body centered cubic lattice for the *high* $T_c$ phase H$_3$S where the electronic and phonons spectrum above $P > 200 GPa$ seem to be consistent in the main and taken as the basis for the further consideration. Discrepancies between results of the theoretical works at *lower* pressure are discussed below.

*Calculating temperature of transition* $T_C$ *in high-* $T_c$ *phase*. The energy scale typical for large bands is few $eV$. At $T = T_C$ the equation for the order parameter $\Delta(\omega_n)$ is:

$$\Delta(\omega_n)Z = -T\sum_m \int d\xi 2\int d\omega \frac{\alpha^2(\omega)F(\omega)}{\omega} D(\omega,\omega_n-\omega_m) \times \frac{\Delta(\omega_m)}{\omega_m^2+\xi^2} \qquad (1)$$

Here $\Delta(\omega_m)/(\omega_m^2+\xi^2)$ is the pairing Greens function [27]; $D(\omega,\omega_n-\omega_m) = -\omega^2/[(\omega_n-\omega_m)^2+\omega^2]$ is the phonon propagator; $\omega$ is the phonon frequency, $\xi$ is the electron energy referred to the chemical potential, $\omega_n = (2n+1)\pi T$ (we employed the method of thermodynamic Green's functions; see, e.g., [28]). The function $\alpha^2(\omega)F(\omega)$ is a well-known quantity determining the strength of the electron-phonon interaction (see, e.g., [29,30]), $F(\omega)$ is the phonon density of states, $Z \simeq 1+\lambda$ in (1) stands for the band mass renormalization. The coupling constant $\lambda$ is defined by the expression:

$$\lambda = 2\int [\alpha^2(\omega)F(\omega)/\omega]d\omega \qquad (2)$$

In [5,6,14-18,20] $T_C$ was calculated substituting $\lambda$ in the McMillan-Dynes-Allen expression[31-33] which depends on $\tilde{\omega}$-the characteristic phonon frequency and $\mu^*$-the Coulomb pseudopotential; usually $\mu^* \approx 0.1 \div 0.15$. Eq. (1) not contains the coupling constant $\lambda$ explicitly. Indeed, the phonon frequency $\omega$ enters not only in the factor $\alpha^2(\omega)F(\omega)$, but in the phonon propagator $D(\omega,\omega_n-\omega_m)$ depending also on $\omega_n-\omega_m$.

In common metals the function $\alpha^2(\omega)F(\omega)$ is characterized by a peak in the phonon density of states (DOS) $F(\omega)$ (see, e.g. [29,34]) that is due to the short-wavelength part of the spectrum where the mode dispersion $\omega(\vec{q})$ is weak. It allows replacing $\omega(\vec{q})$ in the phonon propagator by its average value $\tilde{\omega}$ [30,34,35] (the latter taken either as $<\omega^2>^{1/2}$, see, e.g. [34], or $<\omega_{ln}>$, which is close to $<\omega^2>^{1/2}$, see [33,36]).

The principle cause for concern about the applicability of the same scheme to H$_3$S is that the phonon spectrum of sulfur hydride is complex and consists of the well-separated acoustic and optical branches; the phonon DOS contains *several* peaks so that the introduction of the coupling constant $\lambda$ and the characteristic frequency should be performed with considerable care.

We suggest to separate the phonon spectrum in the two regions of the optical and the acoustic phonons and for each introduce the average frequencies $\tilde{\omega}_{opt}$ and $\tilde{\omega}_{ac}$ and the coupling constants $\lambda_{opt}$ and $\lambda_{ac}$. Eq. (1) accepts the form:

$$\Delta(\omega_n)Z = T\sum_m \int d\xi [(\lambda_{opt}-\mu^*)D(\tilde{\omega}_{opt},\omega_n-\omega_m) + \lambda_{ac}D(\tilde{\omega}_{ac},\omega_n-\omega_m)]\Delta(\omega_m)/(\omega_m^2+\xi^2). \quad (3)$$

Here $\lambda_i = \int_i d\omega \alpha^2(\omega)F(\omega)/\omega$ ; $\tilde{\omega}_i = (2/\lambda_i)\int_i d\omega \alpha^2(\omega)F(\omega)\omega$ , $i \equiv \{opt., ac.\}$ . The critical temperature can be calculated with the use of Eq. (3).

In high-$T_C$ phase (*Im-3m*) *assume* $\lambda_{opt} >> \lambda_{ac}$. Suppose also that $\lambda_{opt} \tilde{<} 1.5$. (These conditions are fulfilled as shown below).

Presenting $T_C$ as $T_C = T_C^0 + \Delta T_C^{ac}$, neglect initially the acoustic phonons and for $T_C^0$ consider the equation containing on the right side of Eq. (3) only the first term. One can use either the McMillan-Dynes expression [31, 32] as the solution for $T_c^0$, or the close expression, obtained analytically in [35] and valid for $\lambda_{opt} \lesssim 1.5$:

$$T_C^0 \approx \frac{\tilde{\omega}_{opt}}{1.2} \exp\left[-\frac{1.04(1+\lambda_{opt})}{\lambda_{opt} - \mu^*(1+0.62\lambda_{opt})}\right]. \qquad (4)$$

To find a correction $\Delta T_C^{ac}$ from the acoustic phonons return to (3). Substituting the total $T_C$ in the first term on the right side of (3) and $T_C^0$ in the second term obtain after calculations:

$$\frac{\Delta T_C}{T_C^0} \approx 2 \frac{\lambda_{ac}}{\lambda_{opt} - \mu^*} \times \frac{\rho^2}{\rho^2 + 1}. \qquad (5)$$

Here $\rho = \tilde{\omega}_{ac} / \pi T_c^0$. These results can be used to evaluate $T_C$ for the high-$T_C$ phase (*Im-3m*).

The coupling constants for the ordinary superconductors are obtained from the tunneling spectroscopy data for $\alpha^2(\omega)F(\omega)$ (see, e.g. [ 30 ]). Such measurements having not been performed for sulfur hydride, we determine the coupling constants $\lambda_{opt}$ and $\lambda_{ac}$ using several theoretically calculated $\alpha^2(\omega)F(\omega)$. Although the results somewhat differ, but they are still relatively close. We use the data [6,13]. Accordingly, $\lambda_{opt} \approx 1.5$ and $\lambda_{ac} \approx 0.5$, in the agreement with above approximations. With these coupling constants and taking for $\tilde{\omega}_{opt}$ and $\tilde{\omega}_{ac}$ the values $\tilde{\omega}_{opt} \approx 1700K$ and $\tilde{\omega}_{ac} \approx 450K$ we obtain $T_C^0 \approx 170K$ and $\Delta T_C^{ac} \approx 45K$, so that the total $T_C \approx 215K$, in a rather good agreement with $T_C \approx 203K$ observed in [4]. As the main contribution comes from the optical phonons, this confirms the self-consistency of our approach.

The fact that the coupling constant $\lambda_{opt}$ in the *Im-3m* phase is so large is a key ingredient determining such a high value of $T_C$. Qualitatively, it is caused by the ability of sulfur to contain several hydrogen atoms in its proximity, that is, by the presence of many light ligands near S atoms.

In the papers mentioned above $T_C$ is calculated without splitting of the phonon spectrum in two parts. The average $\tilde{\omega} = <\omega_{\log}>$ is defined for a whole spectrum, but such approximation is hard to justify (see above). Besides, the McMillan-Dynes equation used to calculate $T_C$ is not valid for such large values of the coupling constant as found in [33-35].

Meanwhile, in the above, $\lambda_{opt}$ is within the limits of the applicability of Eq.(4) allowing to evaluate the relative contribution of the optical and acoustic parts of the phonon spectrum into total $T_C$: ~ 80% of is due to the contribution of the optical phonons and only ~ 20% due to the acoustic part.

*Fine bands structure and role of hole-like pockets.* The fine structure of the electronic energy spectrum in high-$T_C$ phase consists of small hole-like pockets at several spots of the BZ with the Fermi energy of the order $0.5 eV \div 100 meV$. As emphasized above, the feature itself seems to be reliably established in the band calculations [6,9,11,12 (Suppl., Fig.6), 17-19]. (Independently, the existence of small pockets could be confirmed in the tunneling experiments by observation of the two superconductivity gaps. The latter can be analyzed in the framework of the two-gap scenario).

In the literature, however, no agreement exists whether small pockets are essential for achieving so high superconductivity temperature $T_C \approx 203K$ in H$_3$S. As position of a van Hove singularity peak at the Fermi level seems questionable, it is worth considering the possibility of superconductivity arising on a pocket without special suppositions.

Interaction of carriers on small pockets and high frequency phonons cannot be included into the scheme [24], as the Migdal parameter [23] $\omega_{opt}/E_F$ for the hydrogen modes is of the order of unity [17]. Leaving aside the *vH* peak hypotheses [17], temperature $T_C$ for pairing on a pocket can be estimated in the weak-coupling approximation [37].

For simplicity, consider carriers on a single pocket with the Fermi energy $E_F$ interacting with one acoustic mode with the frequency $\omega_{ac} \ll E_F$ and one optical phonon with the frequency $\omega_{ac} \ll \omega_{opt} \sim E_F$. Introduce $2\gamma_{ac}^2 \nu(E_F) = \lambda_{ac}^{pocket}$ and $2\gamma_{opt}^2 \nu(E_F) = \lambda_{opt}^{pocket}$; here $\gamma_{opt}$ and $\gamma_{ac}$ are the matrix elements of the electron-phonon interactions.

In *common* metals dimensionless $\lambda$'s are usually between 1/2 and 1/4. One can assume that $\gamma_{opt}$ and $\gamma_{ac}$ have the magnitude same as in the ordinary metals. Therefore, what makes $\lambda_{ac}^{pocket}$ and $\lambda_{opt}^{pocket}$ small is the differences in DOS compared to the large Fermi surfaces; the $T_C$ value possible on hole-like pockets can be evaluated in the weak coupling limit.

The expression for the temperature $T_C$ of the superconductivity pairing on a pocket acquires the form:

$$T_C = const \times \tilde{\omega}_{ac} \exp[-1/\lambda_T^{pocket}] \times \left(\tilde{\omega}_{opt}/\tilde{\omega}_{ac}\right)^\beta. \quad (6)$$

Here $\tilde{\omega}_{opt} \sim E_F$ and $\beta = \lambda_{opt}^{pocket}/(\lambda_T^{pocket}) \leq 1$. (See in the Supplemental Materials B). Weighing up the uncertainties in DOS $\propto m^* p_F$, taking in (6) $\tilde{\omega}_{ac} \approx 50 meV$ and $\omega_{opt}/\omega_{ac} \approx 3 \div 4$ one admits to $T_c$ between one and a few tens Kelvin.

In the scenario [17] a peak in DOS makes the coupling constants $\lambda_{opt}^{pocket}$ and $\lambda_{ac}^{pocket}$ in Eq. (6) being large enough to account for so high temperature $T_c \approx 180 \div 200K$ in the cubic phase. The superconductivity order first emerging on the pocket, induces the order parameter on large Fermi surfaces.

To repeat, we find this possibility unlikely. A temperature $T_c \approx 215K$ that was obtained above is close to the values estimated for $T_C$ on the *large* Fermi surfaces in [5-8,11-16]. In both cases the transition temperature is correct in magnitude and there is no necessity in additional mechanisms. Besides, as mentioned above, some peak in DOS is usually $0.17 \div 04 eV$ below the chemical potential.

The above estimates for $T_C$ on a pocket further confirm the prevailing role of large Fermi surfaces. We infer, together with [5-8,11-16], that the superconductivity of hybrid sulfur is driven by the photon-mediated pairing on the broad bands. Note in addition that, assuming the leading role of the van Hove singularity DOS peak would result in a change of the prefactor in Eq. (6) $\tilde{\omega}_{ac} \Rightarrow W$ where $W$ is the width of the peak which being of the electronic origin cannot depend on the ionic mass, in stark contradiction with observation of the non-trivial isotope effect [2-4].

*Sequence of phases.* The phase diagram of sulfur hydride was studied in the *ab initio* calculations [5-17] and on few structures have been characterized as the energetically most stable phases. In [6] the *Cccm*-structure realizes below $P<100 GPa$. (For the list of the space groups, see, e. g., Wikipedia). At the pressure $P \geq 200 GPa$ *all* published works agree upon the body centered cubic $Im\bar{3}m$ (*Im-3m*) lattice with one formula $H_3S$ per unit cell. We emphasizethat in this pressure diapason in [5-17] results for the electronic and the phonons spectra differing only in minor details.

At *intermediate* pressure the *first principle* calculations may significantly disagree on the critical pressure and on the symmetry of the phase preceding the *Im-3m* phase. Thus, according to [6], the *Im-3m* phase gives way to the phase *R3m* below $P<180 GPa$. Both in [11] and [6] the *Cccm* structure remains stable up to $P=95 GPa$. In the interval $P=95 \div 150 GPa$ the thermodynamic phase in [11] is *R3m* ($\beta-Po$-type), but the *Im-3m* lattice sets in at the pressure $P=150 GPa$, instead of $\approx 180 GPa$ in [6]. Results for the ground state in [9] are shown only at the two pressures $P=150 GPa$ and $P=200 GPa$. Favorable at $P=200 GPa$ is the *Im-3m* structure, but the *R3m* phase prevails at $P=150 GPa$. The last result contradicts [11], but is in agreement with [6].

*Thermodynamics of the transition.* The $T_C$-growth in the interval of pressure $125 \div 150 GPa$ [3,4] raises a question whether it rapid variation is indeed due to a structural phase transition, and, if true, what are the two adjacent phases. The $T_C$-data in Fig. 3c [4] obtained both while increasing and decreasing the pressure point at the 1$^{st}$ order transition, although the character of a transition cannot be resolved unambiguously from the pressure dependence of $T_C$ alone. As shown above, the accuracy of the *ab initio* calculations is insufficient for determining the precise value of the critical pressure for the transition between the low-$T_C$ and high-$T_C$ phase, although in the analysis of the order of the transition between two phases these uncertainties are less significant than the symmetry arguments. According to [11-13], the transition into the *R3m* phase is driven by the soften sulfur-hydrogen stretching mode. The cubic space group *Im-3m* ($O_h^9$) contains *inversion* among the elements of symmetry. The space group #160 (*R3m*) belongs to the class $C_{3v}$ for which *inversion* is absent. Hence, the second-order transition between the high $T_C$ *Im-3m* phase and the phase *R3m* not contradicts to the Landau theory (see [38]).

This specific result [13] may be sensitive to the calculation details, as for the critical pressure $P_{cr}$ one finds $P=150 GPa$ in [11], while $P_{cr}=103 GPa$ in [13]. However, as one can verify, for the phonon modes at the center of the Brilloiun Zone (BZ) of the point group $O_h$ for the *Im-3m* phase *all four* three-dimensional irreducible representations are *odd* (three vector representations $F_{2u}$ and one $F_{1u}$ from the group $O_h = T_d \times C_i$ [39]), so that any instability at the *gamma*-point would result in the second order character of the transition.

Another possibility is the charge density instability with a non-zero structural vector (one example is the *trigonal R3m (3 f.u./cell)* suggested in [5]). The "imaginary phonon frequencies" in the harmonic approximation in [13] appear in other points of the BZ as well, raising again the accuracy issue. To the best of the authors' understanding, softening of a phonon frequency $\omega(\vec{Q})$ due to its renormalization via the electron-phonon interactions (see [23, 40, 41]) is not discussed in the *first principle* calculations [11-13]. Note in passing the abruptness of the $T_C$-variation [3, 4] that is in favor of the first order transition. To clarify the issue is necessary to perform the X-rays measurements with higher resolution.

*Isotope effect.* The isotopic dependence of $T_C$ (change at the deuterium substitution for hydrogen [2-4] ) is of fundamental importance, since it proves that the high-$T_C$ state is caused by the electron-phonon interaction and it is the high frequency hydrogen modes that determines the value of $T_C$. The optical modes are mainly due to the hydrogen motion, unlike the acoustic modes for which role of the sulfur ions prevails. Indirectly, the value of the isotope coefficient reflects also the relative contributions of each group (optical *vs* acoustic) into the observed $T_C$.

For high-$T_C$ (*Im-3m*) phase the value of the isotope coefficient (in the harmonic approximation)

$$\alpha = -(M/T_C)(\partial T_C/\partial \omega)(\partial \tilde{\omega}_{op}/\partial M) = 0.5(\tilde{\omega}_{op}/T_C)(\partial T_C/\partial \tilde{\omega}_{op}) \qquad (7)$$

can be evaluated from Eqs. (4,5). After the calculation we obtain:

$$\alpha \approx \frac{1}{2}\left[1 - 4\frac{\lambda_{ac}}{\lambda_{opt}} \times \frac{\rho^2}{(\rho^2+1)^2}\right] \qquad (8)$$

(Here $\rho = \tilde{\omega}_{ac}/\pi T_C^0$.). With $\lambda_{opt} \approx 1.5$, $\lambda_{ac} \approx 0.5$, $\tilde{\omega}_{ac} \approx 450K$ (see the Supplemental Material A) we obtain $\alpha \approx 0.35$ in a good agreement with [4]. Note that the value of $\alpha$ can be affected by anharmonicity [12,13] and by the dependence of $\mu^*$ on $\tilde{\omega}_{opt}$, although the last contribution is of the order of $(\mu^*/\lambda_{opt})^2$ and is small.

Notably, the isotope coefficient in the low-$T_C$ phase is different. Indeed, according to [6], the coupling constants there are $\lambda_{opt} \approx \lambda_{ac} \approx 1$ reflecting a larger relative contribution of the acoustic modes. In this case $T_C < \tilde{\omega}_{ac} << \tilde{\omega}_{opt}$ and within the usual BCS logarithmic approximation one can obtain:

$$T_c \approx const \times (\tilde{\omega}_{opt})^{\lambda_{opt}/\lambda_T}(\tilde{\omega}_{ac})^{\lambda_{ac}/\lambda_T}\exp[-(1+\lambda_T)/(\lambda_T-\mu^*)]. \qquad (9)$$

(Compare with Eq. (6): here $\lambda_T = \lambda_{ac} + \lambda_{opt}$; $Z \simeq (1+\lambda_T)$ is included into the exponent [31, 34]). With $\tilde{\omega}_{opt} \approx 105 meV$ and $\tilde{\omega}_{ac} \approx 26 meV$ for the *R3m* in [6] we obtain $T_C \approx 120K$.

From Eqs. (7, 9) $\alpha \approx 0.25$, i. e., noticeable smaller that for high-$T_C$ phase. Experimentally [4], the impact of the isotope substitution in the region of smaller $T_C$ is weaker than in the high-$T_C$ phase in agreement with our analysis.

Smaller $\alpha$ reflects that optical phonons play a larger role in the *Im-3m* phase leading to higher $T_C$.

*Irregular $T_C$-behavior in high-$T_C$ phase and the role of small pockets.* As we argued above, the rapid $T_c$-variation above $P_{cr} \approx 123 GPa$ [4] seems to be better interpreted as the manifestation of the *discontinuous* structural 1st order transition at this pressure. The additional light on the issue is shed by analyzing the subtle contribution of small pockets.

Restricting ourself to the qualitative features, consider a simple two-band model. Let $\Delta(\omega_n)$ and $\Xi(\omega_n)$ be the superconductivity order parameters on the pocket and on the broadband Fermi surface, respectively. Assuming the two bands coupled weakly, the on-pocket interactions change the temperature of transition $T_C^*$ for the whole system only slightly.

Consider, for simplicity only the optical phonons. The linear equation for the parameter $\Xi(\omega_n)$ at $T = T_C^*$ can be written as follows (see in the Supplemental Materials C)

$$\{T_C^* - T_{C0}\}\Xi(\omega_n) = \pi\{\gamma_{12}^2\gamma_{21}^2 / \gamma_{11}^2\}\nu_P(E_F)T\sum_m |D_{opt}(\omega_n - \omega_m)|(1/|\omega_m|)\Xi(\omega_n) \quad (10)$$

In (10) $|D_{opt}(\omega_n - \omega_m)| = \omega_{opt}^2 / [(\omega_n - \omega_m)^2 + \omega_{opt}^2]$, $\gamma_{11}$ and $\gamma_{12}$ are the matrix elements of electron-phonon interaction on the large Fermi surface and that for scattering of an electron between the large and the small Fermi surfaces, correspondingly ($\gamma_{12} << \gamma_{11}$). (The critical temperature $T_0^* > T_{C0}$).

The density of states on the large Fermi surface $N_L(E_F^L) \sim m_{eLFS} p_{LFS} / (2\pi)^2$ exceeds that one on the pocket $N_P(E_F^P) = m_{eP} p_{FP} / (2\pi)^2$. Therefore the change in the temperature of transition $T_C^* - T_{C0}$ as a function of pressure is just proportional to DOS on the pocket. Assume the 1st order transition takes place at $P_{cr} \approx 123 GPa$. $T_c$ changes from $T_c \approx 100K$ to $T_c \approx 200K$ [2-4] with the pockets emerging simultaneously with the onset of the cubic *Im-3m* phase. *A decrease* of $T_c^*$ after the high $T_C$ phase onset, according to (10), corresponds to the pocket size $p_{FP}$ *shrinking* with applying higher pressure. This interpretation is in contrast with the scenario [17] of the pockets developing via the Lifshitz 2.5-topological transition, for in that case the pockets sizes would *grow* with pressure.

*In summary*, having scrutinized results of the *ab-initio* calculations, we concluded that the accuracy of the state-of-art *first principle* methods is not sufficient to identify the character of the thermodynamic transition between the high- and low-$T_C$ phase of $H_3S$ unambiguously. We gave arguments for the 1st order phase transition, possibly related to a CDW instability, is a most credible hypothesis that accounts for a step-like increase of $T_c$ at $P_{cr} \approx 123 GPa$ [4] responsible for so considerable increase of $T_C$ from $\approx 100K$ in the low-$T_c$ phase up to $\approx 200K$ in the high-$T_c$ phase attributed to the contribution into the superconductivity pairing of high frequency hydrogen modes prevailing over the acoustic modes. In the

low-$T_c$ phase the two phonons groups contribute into $T_C$ equally. We demonstrate that the $T_c$-decrease in high-$T_c$ phase above $T_{C\max}$ reflects the emergence of small hole-like pockets at the 1st order transition. Our analysis shows that the methods of calculating $T_C$ based on the McMillan extrapolation, successful in the ordinary superconductors, are not applicable to $H_3S$ because of its complex phonons spectrum comprised of the acoustic and several optical hydrogen modes with much higher frequency. The proposed modification for calculating pairing on large Fermi surfaces provides the realistic values for the temperature of the superconductivity onset. The calculated isotopic dependence of $T_C$ turns out be different on the two sides of the transition, in agreement with [3, 4].

Comparing the contributions into the $T_C$-value from large Fermi surfaces and from a pocket we infer that superconductivity in $H_3S$ is driven by interactions on the former. We commented that the presence of small pockets in high $T_C$-phase can be revealed by observation in the tunneling experiments of *two* superconducting gaps in the energy spectrum of $H_3S$ at low temperatures.

## Acknowledgments

The authors thank M.I. Eremets for clarification of a number of significant experimental details and providing us with more useful literature references and to M. Calandra for stimulating discussions and sharing some of his theoretical results before publication. The work of LPG is supported by the National High Magnetic Field Laboratory through NSF Grant No. DMR-1157490, the State of Florida and the U.S. Department of Energy.

# Supplementary Materials of

*Pressure and high $T_C$ superconductivity: applications to sulfur hydrides*

Lev P. Gor'kov and V. Z. Kresin

### A. Calculation of critical temperature

In Eq. (3) of the text integrate over $\xi$:

$$\Delta(\omega_n)Z = \pi T \sum_m [\lambda_{opt} D(\tilde{\omega}_{opt}, \omega_n - \omega_m) + \lambda_{ac} D(\tilde{\omega}_{ac}, \omega_n - \omega_m)]\Delta(\omega_m)/|\omega_m|.$$

Assume that the main contribution to the critical temperature in the high-$T_C$ phase is due to optical phonons $\lambda_{opt} \gg \lambda_{ac}$ ( $\lambda_{opt} \lesssim 1.5$, see below). Let $T_C \approx T_C^0 + \Delta T_C^{ac}$. For $T_C^0$ use of the McMillan-Dynes expression:

$$T_c^0 \approx \frac{\tilde{\omega}_{opt}}{1.2} \exp[-\frac{1.04(1+\lambda_{opt})}{\lambda_{opt} - \mu^*(1+0.62\lambda_{opt})}]. \quad (A.1)$$

(This expression valid for $\lambda_{opt} \lesssim 1.5$ [33,35] was obtained by including the contribution $Z \simeq 1+\lambda$ and the fitting to $T_C$ for Nb (see, e. g., the discussion in [34]). It appears to be a good description for other strong-­-coupled superconductors.

One can use also the expression, obtained analytically in [24] and valid for $\lambda_{opt} \lesssim 1.5$:

$$T_C \approx 1.14\tilde{\omega}_{opt} \exp\left[-\frac{1+1.5\lambda_{opt}}{\lambda_{opt} - (1+0.5\lambda_{opt})}\right] \quad (A.\ 1')$$

Substituting $T_C \approx T_C^0 + \Delta T_C$ into Eq. (3) one finds $\Delta T_C \equiv \Delta T_C^{ac}$-the correction due to the acoustic mode. At that one can assume [35] $\Delta(\omega_m) = \Delta$. In the equation:

$$\pi T \sum_m [\lambda_{opt}\{-\frac{2\Delta T}{T_0} \times \frac{\tilde{\omega}_{opt}^2(\omega_n-\omega_m)^2}{[\tilde{\omega}_{opt}^2+(\omega_n-\omega_m)^2]^2}\} + \lambda_{ac}\frac{\tilde{\omega}_{ac}^2}{[\tilde{\omega}_{ac}^2+(\omega_n-\omega_m)^2]}]/|\omega_m| = 0 \quad (A.2)$$

omit the term $\omega_n$:

$$\pi T \sum_m [\lambda_{opt}\{-\frac{2\Delta T}{T_0} \times \frac{\tilde{\omega}_{opt}^2(\omega_n-\omega_m)^2}{[\tilde{\omega}_{opt}^2+(\omega_n-\omega_m)^2]^2}\}/|\omega_m| \Rightarrow -\lambda_{opt}\frac{2\Delta T}{T_0}\pi T \sum_m \frac{\tilde{\omega}_{opt}^2|\omega_m|}{[\tilde{\omega}_{opt}^2+(\omega_m)^2]^2}.$$

In high-$T_c$ phase $2\pi T_0 \approx 1200K \leq \omega_{opt}$; for an estimate present the sum over $m$ as the integral

$$-\lambda_{opt}\frac{2\Delta T}{T_0}\pi T \sum_m [\frac{\tilde{\omega}_{opt}^2|\omega_m|}{[\tilde{\omega}_{opt}^2+(\omega_m)^2]^2}\} \Rightarrow -\lambda_{opt}\frac{\Delta T}{T_0}\int\frac{\tilde{\omega}_{opt}^2 \omega d\omega}{[\tilde{\omega}_{opt}^2+\omega^2]^2} = -\lambda_{opt}\frac{\Delta T}{T_0}.$$

In the second term in (A.2) contribution from $\omega_n = \omega_m$ compensates $\lambda_{opt}$ in $Z \simeq 1+\lambda_{opt}$. As $2\pi T_0 \approx 2.7\tilde{\omega}_{ac}$, leaving only the next two terms obtain:

$$\Delta T_C \approx 2T_C^0 \frac{\lambda_{ac}}{\lambda_{opt}-\mu^*} \times \frac{\rho^2}{\rho^2+1} \quad (A.3)$$

Here $\rho = \tilde{\omega}_{ac}/\pi T_C^0$. We use $\alpha^2(\omega)F(\omega)$ from [5-9] to determine $\lambda_{opt}$ and $\lambda_{ac}$; from [5] $\lambda_{opt} \approx 1.5$ and $\lambda_{ac} \approx 0.5$. Taking $\tilde{\omega}_{op} \approx 1700K$ and $\tilde{\omega}_{ac} \approx 450K$ we obtain $T_C^0 \approx 170K$ and $\Delta T_C^{ac} \approx 45K$, i. e., $T_C \approx 215K$. The estimate is in a rather good agreement with $T_C \approx 203K$, ($\lambda_{opt}$ and $\lambda_{ac}$ can be measured, at least in principle, with the help of the tunneling spectroscopy (see, e.g. [29, 30])).

**B. Estimates of temperature of transition on a pocket in the weak –coupling limit**

The equation for the pairing order parameter on the pocket $\Delta(\omega_n)$ at $T=T_C$ is:

$$\Delta(\omega_n) = -T\sum_{m>0}\int 2\nu(E_F)d\xi\Delta(\omega_m)[(\omega_m^2+\xi^2)]^{-1}\times\left[\gamma_{ac}^2 D_{ac}(\omega_{ac};\omega_n-\omega_m)+\gamma_{opt}^2 D_{opt}(\omega_{opt};\omega_n-\omega_m)\right]. \quad (B.1)$$

(All notations are as in Eq. (4) of the text). In (B.1) $\nu(E_F)$ is the density of states (DOS) on the pocket Dispersion of the acoustic modes is omitted for brevity. The critical temperature $T_c$ on the *isolated* pocket is estimated in the logarithmic approximation assuming small $2\gamma_{ac}^2\nu(E_F)=\lambda_{ac}$ and $2\gamma_{opt}^2\nu(E_F)=\lambda_{opt}$.

Integration and summation in Eq. (B.1) for contributions from the acoustic and the optical modes are limited by the phonons propagators and $E_F$ for the characteristic band energy. Introduce $\Delta_1$ and $\Delta_2$ for average values of the superconducting order parameter $\Delta(\omega_n)$ at $\omega_n<\omega_{ac}$ and $\omega_{ac}<\omega_n<\omega_{opt}\sim E_F$. Eq. (B.1) reduces to the algebraic system:

$$\Delta_1 = 2\gamma_{ac}^2\nu(E_F)\ln(\omega_{ac}/T_C)\Delta_1 + 2\gamma_{opt}^2\nu(E_F)[\ln(\omega_{opt}/T_C)\Delta_1+\ln(\omega_{opt}/\omega_{ac})\Delta_2]$$
$$\Delta_2 = 2\gamma_{opt}^2\nu(E_F)[\ln(\omega_{opt}/T_C)\Delta_1+\ln(\omega_{opt}/\omega_{ac})\Delta_2]$$

The temperature for superconductivity transition $T_c$ on the *isolated* pocket ($\lambda_{opt}\sim\lambda_{ac}\ll 1$ and small $\lambda_{opt}\ln(\omega_{opt}/\omega_{ac})$) is:

$$T_C = const\times\omega_{ac}\exp[-\frac{1}{\lambda_{ac}+\lambda_{opt}}]\times\left(\frac{\omega_{opt}}{\omega_{ac}}\right)^\beta. \quad (B.2)$$

Here $\omega_{opt}\sim E_F$ and:

$$\beta = \lambda_{opt}/(\lambda_{ac}+\lambda_{opt})\leq 1. \quad (B.3)$$

## C. Two band problem

Let the common approach be employed for the pairing parameter $\Xi(\omega_n)$ on the large Fermi surface. For simplicity, consider interaction with one optical mode. Weak coupling to the pocket means a *perturbation* of $T_c$ for the whole system. After simple transformations the resulting equation for the parameter $\Xi(\omega_n)$ in Eq. (B.1) reads:

$$\Xi(\omega_n) = \pi T\sum_m \gamma_{11}^2 |\nu_L(E_F)| D_{opt}(\omega_n-\omega_m)|(1/|\omega_m|)\Xi(\omega_n)$$

$$+\pi T\sum_k \gamma_{12}^2 |\nu_P(E_F)| D_{opt}(\omega_n-\omega_k)|(1/|\omega_k|)\times\pi T\sum_m\gamma_{21}^2|\nu_L(E_F)|D_{opt}(\omega_k-\omega_m)|(1/|\omega_m|)\Xi(\omega_m).$$
(C.1)

In (C.1) $|D_{opt}(\omega_n-\omega_m)|=\omega_{opt}^2/[(\omega_n-\omega_m)^2+\omega_{opt}^2]$, $\gamma_{11}$ stands for the matrix elements for electron-phonon interaction on the large Fermi surface and $\gamma_{12}$-for the scattering between the large Fermi surface and small pocket ($\gamma_{12}\ll\gamma_{11}$). Leaving the first term on the right hand side Eq. (C.1) one defines $T_{C0}$- temperature of transition for the large Fermi surface. The second term is the additional contribution from the pocket. At $|T-T_{C0}|<<T_{C0}$ one can rewrite (C.1) as:

$$\{T_C^* - T_{C0}\}\Xi(\omega_n) = \pi\{\gamma_{12}^2\gamma_{21}^2/\gamma_{11}^2\}v_P(E_F)T\sum_m |D_{opt}(\omega_n - \omega_m)|(1/|\omega_m|)\Xi(\omega_n). \quad (C.2)$$